\documentclass[aps,prb,twocolumn,showpacs,superscriptaddress,groupedaddress]{revtex4-1}
\usepackage{amsmath}    
\usepackage{graphicx}   
\usepackage{verbatim}   
\usepackage{color}      
\usepackage{subfigure}  

\begin{document}
\title{Weak Antilocalization and Conductance Fluctuation in 
a Sub-micrometer-sized Wire of Epitaxial Bi$_{2}$Se$_{3}$}

\author{Sadashige Matsuo, Tomohiro Koyama, Kazutoshi Shimamura, Tomonori
Arakawa, Yoshitaka Nishihara, Daichi Chiba, Kensuke Kobayashi}
\email{kensuke@scl.kyoto-u.ac.jp} \author{Teruo Ono} \affiliation{Institute for
Chemical Research, Kyoto University, Uji, Kyoto 611-0011, Japan}

\author{Cui-Zu Chang, Ke He, Xu-Cun Ma, Qi-Kun Xue} \affiliation{Beijing
National Laboratory for Condensed Matter Physics, Institute of Physics, Chinese
Academy of Sciences, Beijing 100190, China}

\begin{abstract}
In this paper, we address the phase coherent transport in a sub-micrometer-sized
Hall bar made of epitaxial Bi$_{2}$Se$_{3}$ thin film by probing the weak
antilocalization (WAL) and the magnetoresistance fluctuation below 22~K.  The
WAL effect is well described by the Hikami-Larkin-Nagaoka model, where the
temperature dependence of the coherence length indicates that electron
conduction occurs quasi-one-dimensionally in the narrow Hall bar.  The
temperature-dependent magnetoresistance fluctuation is analyzed in terms of the
universal conductance fluctuation, which gives a coherence length consistent
with that derived from the WAL effect.
\end{abstract}

\maketitle
\section{Introduction}
A topological insulator (TI) is a new phase of matter that was theoretically
predicted~\cite{Kaneprl2005,Bernevigscience2006,Fuprb2007} and was later
realized in two-dimensional~\cite{Konigscience2007} and
three-dimensional~\cite{Hsiehnature2008} materials.  This new phase originates
from the strong spin-orbit interaction to cause band inversion. As a result, TI
has a band gap in the bulk, while there exist gapless states on the
surface. Two-dimensional TI has one-dimensional states at the edge and
three-dimensional TI (3DTI) has two-dimensional states at the surface.
Electrons in these states satisfy the linear dispersion relation and their spins
are polarized.  In particular, in 3DTI, electrons in the surface states behave
as Dirac electrons~\cite{hasanrmp2010}.  These gapless surface states are
expected to be robustly protected against the time-reversal-invariant
perturbation, and therefore, it is expected that they can be applied to
spintronics ~\cite{Yazyevprl2010} and quantum computing~\cite{Akhmerovprl2009}.

In order to address and make the best use of this protection as well as the
electronic properties as Dirac electrons, it is necessary to experimentally
investigate TI from the viewpoint of quantum transport.  Some quantum transport
phenomena in 3DTI have already been observed; several groups reported the weak
antilocalization (WAL) effect in 3DTI
crystals~\cite{checkelskyprl2009,Checkelskyprl2011},
nanoribbons~\cite{quprl2011}, and thin films~\cite{chenprb2011,wangprb2011,
Steinbergprb2011,Chenprl2010, Heprl2011,Liuprb2011,Kimprb2011,onoseapex2011}.
Considerable fluctuation in the magnetoresistance, similar to the universal
conductance fluctuation (UCF), was also reported for bulk
crystals~\cite{checkelskyprl2009}.  Nevertheless, only a few studies have
simultaneously addressed and analyzed both the WAL effect and the
magnetoresistance fluctuation.


Bi$_{2}$Se$_{3}$ is one of the representative 3DTIs with a single Dirac cone; it
was predicted~\cite{zhangnatphys2009} and established by angle-resolved
photoemission spectroscopy (ARPES)~\cite{xianatphys2009, hsiehnature2009} in
2009. Because the bulk band gap of Bi$_{2}$Se$_{3}$ is relatively large compared
to that of other 3DTI materials, it has attracted attention from the viewpoint
of 3DTI and the Landau levels of surface states~\cite{chengprl2010,
Hanaguriprb2010}, and its thickness dependence~\cite{Zhangnatphys2010,
Kimprb2011} has been reported. However, two technical difficulties were
encountered in these studies. First, the synthesized Bi$_{2}$Se$_{3}$ material
is often naturally doped by electrons because of Se
vacancies~\cite{xianatphys2009}, and therefore, the bulk states could contribute
to the transport properties of the Bi$_{2}$Se$_{3}$ sample, which prevents the
pure surface state contribution from being determined.  Second, the surface of
Bi$_{2}$Se$_{3}$ is sensitive to water and oxidation, and therefore, the
electron mobility in this state is usually
small~\cite{butchprb2010,analytisprb2010}.

\begin{figure}[t]
\begin{center}
\includegraphics[width=.95\linewidth]{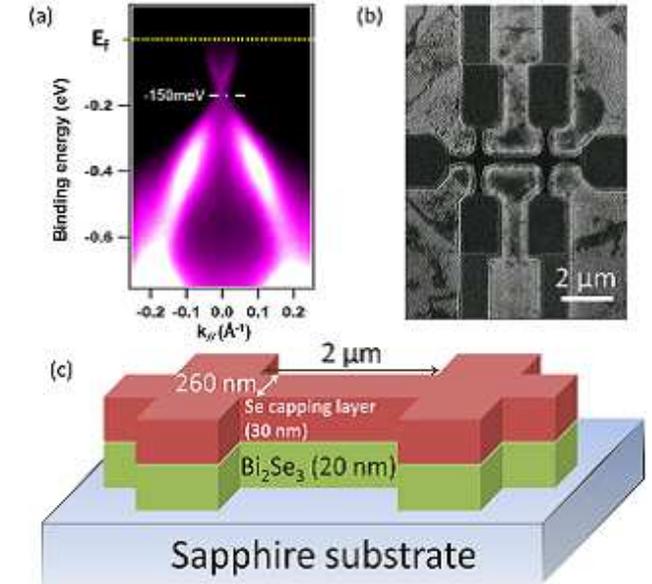}
\end{center}
\vskip-\lastskip \caption{(a) ARPES spectra of epitaxially grown 20-nm-thick
film of Bi$_{2}$Se$_{3}$.  (b) SEM image of the present Hall bar sample.  The
scale (white bar) is $2$ $\rm{\mu}$m. (c) Schematic of geometry of the narrow
Hall bar sample.  A protective amorphous Se layer was grown on 20 quintuple
Bi$_{2}$Se$_{3}$ layers.  We fabricated the Hall bar geometry from the thin film
by using EBL and Ar ion milling. } \label{fig:1}
\end{figure}

Herein, we report the WAL effect and magnetoresistance fluctuation in a
sub-micrometer-sized Hall bar (wire) sample fabricated on an epitaxial
Bi$_{2}$Se$_{3}$ thin film.  We use Bi$_{2}$Se$_{3}$ thin film to minimize the
bulk contribution and to address the mesoscopic coherence by using the WAL
effect and the magnetoresistance fluctuation at low temperatures. We show that
the WAL effect is well described by the Hikami-Larkin-Nagaoka model and that the
temperature dependence of the coherence length indicates that electron
conduction occurs quasi-one-dimensionally in the narrow Hall bar.  The
temperature-dependent magnetoresistance fluctuation is analyzed in terms of UCF,
which gives a coherence length consistent with that derived from the WAL effect.

\section{Experiment}
We grew a 20-nm-thick Bi$_{2}$Se$_{3}$ thin film by molecular beam epitaxy (MBE)
on a sapphire substrate, as described previously~\cite{changspin2011}. This
thickness is equivalent to 20 quintuple layers.  The ARPES measurement clarified
the electron band structure in the momentum-energy space under the Fermi energy,
as shown in Fig.~\ref{fig:1}(a).  The Fermi energy is located 150 meV above the
Dirac point and in the bulk conduction band.  This is due to the natural doping
by Se vacancies.  Therefore, the carriers of both surface states and the bulk
conduction band contribute to the transport phenomena of the thin films.

We deposited a 30-nm-thick amorphous Se layer to prevent Bi$_{2}$Se$_{3}$ from
being destroyed by water or oxidation.  The thin film was fabricated into a
narrow 260-nm-wide wire-shaped Hall bar by using electron beam lithography (EBL)
and Ar ion milling.  Ti (5~nm) and Au (100 nm) were then deposited as
electrodes.  Figure~\ref{fig:1}(b) shows a scanning electron microscope (SEM)
image of the Hall bar sample.  A schematic of the sample geometry is shown in
Fig.~\ref{fig:1}(c). We measured the magnetic-field dependence of the
longitudinal resistance and the Hall resistance at 11 different temperatures
between 2 and 22~K.  All the experiments were carried out by using the
standard lock-in technique. We measured two devices fabricated
from the same Bi$_{2}$Se$_{3}$ film and obtained consistent results as reported
below.

\section{Results and Discussions}
\subsection{Basic parameters}
\begin{figure}[t]
   \begin{center}
   \includegraphics[width=.95\linewidth]{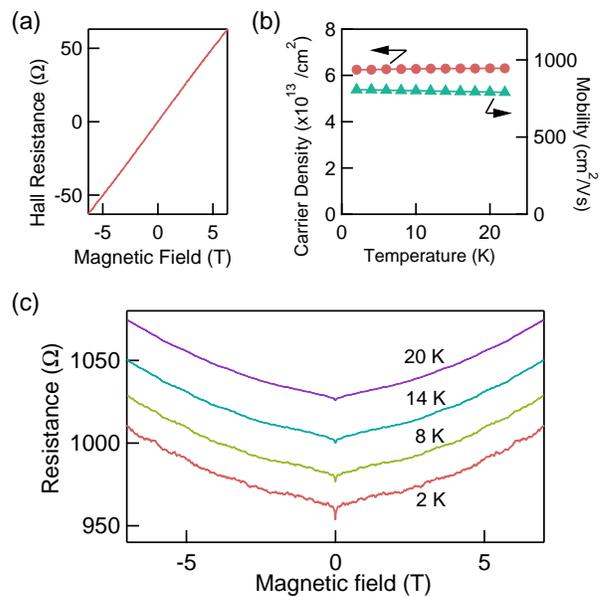}
   \end{center}
\vskip-\lastskip \caption{(a) Hall resistance at 2 K as a function of
the magnetic field. (b) The carrier density and mobility are shown as a function of
temperature; they do not depend on the temperature.  (c) The longitudinal resistance at
2, 8, 14, and 20 K is shown.  The $y$-axis shows the data obtained at 2~K;
the other data are incrementally shifted upward by 21 $\Omega$ for clarity.  There
are dips near 0~T in all the traces. In addition, the resistance fluctuates as
the magnetic field changes.  The fluctuation increases as the temperature decreases.}
\label{fig:2}
\end{figure}

First, we present the basic electronic properties of our device.
Figure~\ref{fig:2}(a) shows the Hall resistance as a function of magnetic field
between -7 and 7~T at 2~K.  The carriers are electrons,the density of which is
$n_s=6.2\times 10^{13} \rm{/cm^2}$.  The electron mobility, $\mu = l/(weR n_s)=
807$ $\rm{cm^2/Vs}$, and the mean free path of electrons, $l_m = \hbar \mu
\sqrt{2\pi n_s}/e = 105$~nm, are obtained at 2~K based on the longitudinal
resistance $R$ at 0 T and the length, $l=2.0$~$\rm{\mu m}$, and width,
$w=260$~nm, of the Hall bar~\cite{beenakkerssp1991}.  The carrier density and
mobility are shown as a function of temperature in Fig.~\ref{fig:2}(b).  Both
show little dependence on temperature, indicating that our sample is in the
metallic regime.  

When the number of electron bands that contribute to the
carrier transport is more than one, the Hall resistance should have nonlinear
components as a function of the magnetic field. Actually, such a nonlinearlity
was reported for the TI devices already~\cite{Analytisnatphys2010}. In our case,
however, the applied magnetic field ($\lesssim7$~T) was not high enough to
observe the expected nonlinearity to prove that both the surface and bulk states
contribute to transport. Thus, the present data obtained in the transport
measurement are not sufficient to affirm that the surface states play a role in
the conduction, while the ARPES result in Fig.~\ref{fig:1}(a) clearly implies
that there are both surface states and bulk conduction states at the Fermi
level.  Based on this fact, it is believed that electrical current flows through
both the surface states and the bulk conduction states. We will discuss this
point later based on our experimental findings.

Two important features are noted in the longitudinal resistance as a function of
the magnetic field (Fig.~\ref{fig:2}(c)).  One is a dip structure around 0 T,
which originates from the WAL effect.  The other is the resistance fluctuation
occurring over a wide magnetic field range, which becomes prominent at lower
temperatures and is thus similar to UCF. These two features are the main topics
of this paper, as discussed in the following.

\subsection{Weak antilocalization}
\begin{figure}[t]
\begin{center}
\includegraphics[width=.95\linewidth]{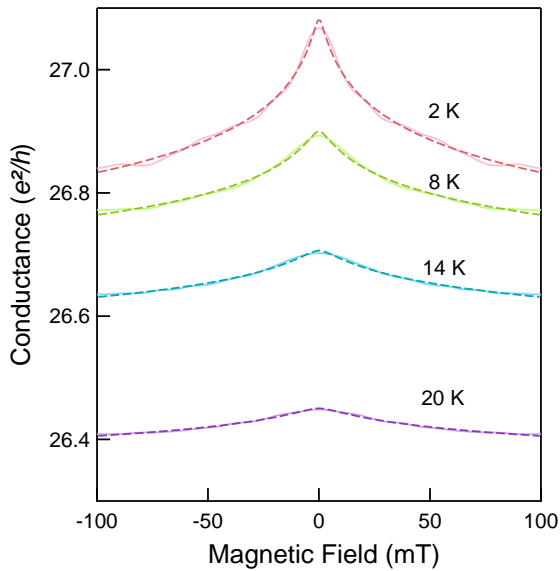}
\end{center}
\vskip-\lastskip \caption{Magnetoconductance around 0 mT is shown.  Solid
curves indicate the experimental results and dashed curves indicate the results of
the fitting by the HLN model.  The $y$-axis shows the data obtained at 2 K; the
other data are incrementally shifted downward by $0.12e^2/h$. } \label{fig:3}
\end{figure}

We first consider the WAL effect. Without spin-orbit interaction, quantum
interference between the time reversal pair of electron waves scattered by
impurities is destructive and results in electron localization. This
localization gives rise to the conductance correction to reduce the conductance
around a zero magnetic field.  This quantum interference occurs at the scale of
the coherence length. If there is strong spin-orbit interaction, on the
contrary, the correction enhances conductance, which is known as the WAL effect.
Actually, as seen in Fig.~\ref{fig:2}(c), the dip structure in the resistance
becomes more remarkable with decreasing temperatures. The data shown in
Fig.~\ref{fig:3} are the conductance around zero magnetic fields [converted from
the data in Fig.~\ref{fig:2}(c)].  The enhancement in the peak structure as the
temperature decreases is characteristic of the WAL effect.

\begin{figure}[t]
\begin{center}
\includegraphics[width=.95\linewidth]{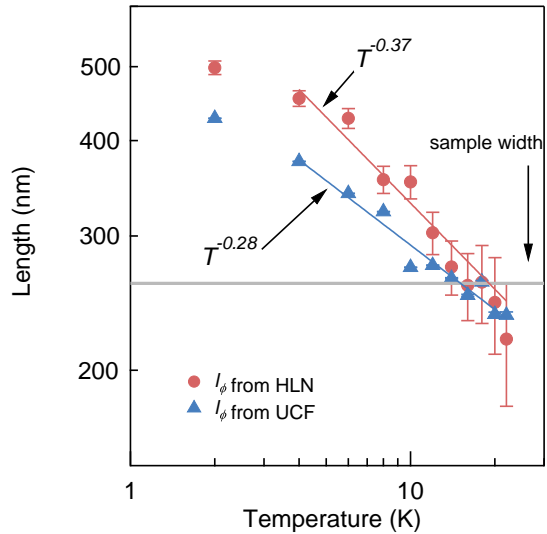}
\end{center}
\vskip-\lastskip \caption{Coherence lengths obtained by two different methods
are shown as a function of temperature.  Circles and triangles
respectively indicate that the coherence length is derived from the WAL effect
using the HLN model and conductance
fluctuation using the UCF theory.  The horizontal line indicates the width of
the Hall bar sample ($w=260$~nm).  The coherence length derived from the WAL
effect based on the HLN model is consistent with that derived from the UCF. The
two solid lines show the results of the fitting with a power-law function of
temperature. } \label{fig:4}
\end{figure}

We first assume that the electrons in the present device behave as in a
two-dimensional system and analyze the magnetoconductance with two-dimensional
WAL models. In the case of a two-dimensional system, when a magnetic field is
applied perpendicularly to the two-dimensional plane, the conductance correction
gradually vanishes with an increase in the magnetic field because of
the breaking of the time-reversal symmetry.  Therefore, the effect of the
quantum interference weakens. Conventionally, this reduction in conductance can
be explained by the Hikami-Larkin-Nagaoka (HLN) model~\cite{HLN}, where the spin relaxation is governed by the Elliot-Yafet mechanism~\cite{EYE,EYY}.  In this
model, the conductance correction $\delta G_{\rm{WAL}}(B)$ is given by
\begin{eqnarray}
\delta G_{\rm{WAL}}(B) &\equiv & G(B)-G(0) \nonumber \\
&=& \alpha \frac{e^2}{2\pi ^2\hbar}[\psi (\frac{1}{2} + \frac{B_\phi }{B})
 -\ln (\frac{B_\phi }{B})]. \label{eq:1}
\end{eqnarray}
$G(B)$ is the conductance as a function of magnetic field, $B$. $\psi$ represents
the digamma function.  $B_\phi = \hbar/4el_\phi^2$ is a magnetic field
characterized by coherence length $l_\phi$.  In the case of a system with
strong spin-orbit interaction, the prefactor $\alpha$ is equal to $-0.5$ and
this model explains the conductance correction of the WAL effect.

The HLN model has already been applied to explain the results of the WAL effect
in Bi$_{2}$Se$_{3}$
samples~\cite{quprl2011,chenprb2011,wangprb2011,Steinbergprb2011}.  Following
these studies, we analyzed the experimental results of magnetoconductance using
the HLN model and estimated the coherence length at each temperature. To
estimate the prefactor and coherence length, we fitted the magnetoconductance
with Eq.~(\ref{eq:1}), as shown in Fig.~\ref{fig:3}.  Equation~(\ref{eq:1}) can be used
to describe the conductance correction when the magnetic field is less than $B_m
= \hbar/2el_m^2$.  $B_m$ may be derived from the mean free path of our sample to
be $\sim~120$~mT so that we can use the conductance peak within $\pm 100 $ mT
for the fitting.  The prefactor $\alpha $ is approximately $-0.3$ at 2 K, which
is of the same order as in the symplectic case ($\alpha = -0.5$).  This means
that carrier electrons are strongly affected by strong spin-orbit interaction.
The obtained coherence lengths as a function of temperature are shown in
Fig.~\ref{fig:4}.  The coherence length increases from 0.2~$\mu$m to around
0.5~$\mu$m as the temperature decreases from 22 to 2 K.  The observed behavior
of the coherence length is consistent with that reported for thin films of
Bi$_{2}$Se$_{3}$~\cite{onoseapex2011,wangprb2011}.

From the temperature dependence of the coherence length, the dimensionality of
the system is known. Theoretically, the coherence length is proportional to
$T^{-1/2}$ for the two-dimensional system and $T^{-1/3}$ for the one-dimensional
one~\cite{altshulerjpc1982}.  Figure~\ref{fig:4} shows that the coherence length
is proportional to $T^{-0.37}$ above $\sim 4$ K~\cite{attention1}, which is very
close to the exponent expected for the one-dimensional system.  Remarkably, as
shown in Fig.~\ref{fig:4}, $l_\phi>w$ at $T<15$ K, which is consistent with the
fact that electrons move quasi-one-dimensionally. Although the above analysis is
made based on the two-dimensional case, as a consistency check, the
one-dimensional model is used to analyze the observed WAL
effect~\cite{kurdakprb1992}. We have confirmed that the coherence lengths based
on this one-dimensional model are very consistent with what has been derived
from the conventional HLN model.

\subsection{Magnetoresistance fluctuation}
\begin{figure}[t]
   \begin{center}
   \includegraphics[width=.95\linewidth]{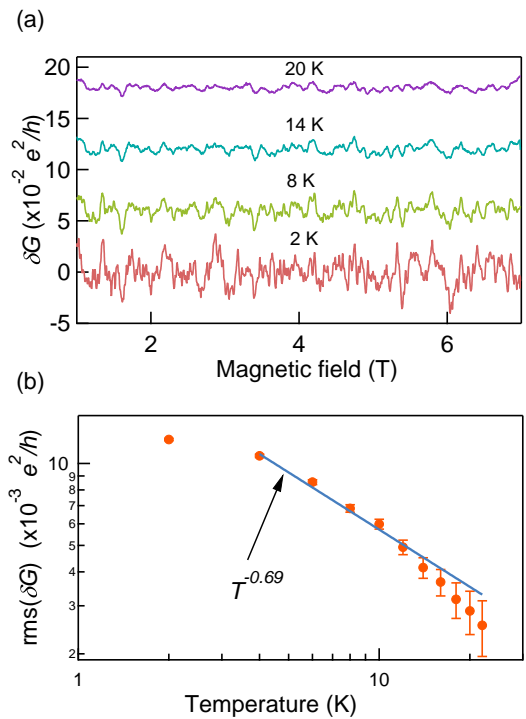}
   \end{center}
\vskip-\lastskip \caption{(a) Fluctuation components of magnetoconductance. The
curves represent the fluctuation at 2, 8, 14, and 20 K.  The $y$-axis shows the
data for 2 K; the other data are incrementally shifted upward by 0.06$e^2/h$.
(b) Root mean square of the conductance fluctuation as a function of
temperature. The value increases as the temperature decreases. The temperature
dependence of the conductance fluctuation from 4 to 22 K is fitted to be
$T^{-0.69}$, as shown by the straight line.}  \label{fig:6}
\end{figure}

As shown in Fig.~\ref{fig:2}(c), the magnetoresistance fluctuates as the
magnetic field changes. We observe that this fluctuation is perfectly
reproducible at a fixed temperature and that it becomes prominent as the
temperature lowers. Now we show that this can be attributed to UCF, a typical
mesoscopic effect. The conductance of mesoscopic samples differs from each
other, and if phase coherence is maintained over the entire sample, the
conductance fluctuation $\delta G$ is nearly equal to a universal value, $e^2/h$
at 0 K~\cite{altshulerjetplett1985,leeprl1985}.  We calculated the correlation
function $F(\Delta B)$ of the components of the conductance fluctuation $\delta
G(B)$ by extracting the smoothed conductance from the original conductance data
between 1 and 7 T, as shown in Fig.~\ref{fig:6}(a). The
fluctuations $\delta G(B)$ in Fig.~\ref{fig:6}(a) are reproducible as a function of
magnetic field at different temperatures, while their amplitude decreases as the temperature increases.  This supports that the conductance fluctuation is
intrinsic to the device transport. The correlation function is defined by
$F(\Delta B) \equiv \langle \delta G(B+\Delta B)\delta G(B)\rangle$, where
$\langle ...\rangle$ expresses the ensemble average.

First, we estimated the coherence length from the obtained correlation function.
In the UCF theory, the conductance fluctuation $\delta G(B)$ correlates with
$\delta G(B+B_c)$ when $B<B_c$, where $B_c$ is characterized by the coherence
length.  As the coherence length derived from the WAL effect analyzed using the
HLN model shows the temperature dependence expected for a one-dimensional
system, we assumed that the present sample is one-dimensional and analyzed the
conductance fluctuation as one-dimensional UCF.  $B_c$, defined by
$F(B_c)=F(0)/2$, is expressed as $B_c= 0.95h/ew{l_\phi}$ in a one-dimensional
system that is in the diffusive regime with $l_\phi > l_T$. Here, $l_T$ is a
thermal diffusion length~\cite{beenakkerprb1988}.  As shown in Fig.~\ref{fig:4},
the coherence length derived based on UCF is satisfactorily consistent with that
derived from the WAL effect.  These coherence lengths are proportional to
$T^{-0.28}$, which is consistent with the quasi-one-dimensional case. The
observed quasi-one-dimensionality is also consistent with the results from the
WAL effect using the HLN model.

Second, we investigated the temperature-dependent behavior of the root mean
square of the conductance fluctuation defined by rms$(\delta G) \equiv
{F(0)}^{1/2}$.  According to the UCF theory~\cite{leeprb1987} for the
one-dimensional system, if $l>l_\phi$ holds, rms$(\delta G)$ reflects the size
of the sample, which causes rms$(\delta G) \propto (l_\phi/l)^{3/2}$. In
addition, if $l_\phi>l_T$ holds, rms$(\delta G)$ is affected more greatly by the
finite temperature, which yields rms$(\delta G) \propto l_T/l_\phi$. Therefore,
rms$(\delta G) \propto l_T l_\phi^{1/2}/l^{2/3}$ is expected.  As already
discussed, $l_\phi \propto T^{-1/3}$ and $l_T \propto T^{-1/2}$; therefore,
rms$(\delta G) \propto T^{-2/3}$.  In order to confirm this, we investigated the
temperature dependence of rms$(\delta G)$.  Figure~\ref{fig:6}(b) shows
rms$(\delta G)$ as a function of temperature.  The temperature dependence of
rms$(\delta G)$ was fitted by $T^{-0.69}$ except for the data at 2
K~\cite{attention1}, which is consistent with the prediction of the UCF
theory~\cite{leeprb1987}, $T^{-2/3}$.  This suggests that the observed
fluctuation originates from UCF in a quasi-one-dimensional system.

Recently, a similar magnetoresistance fluctuation signal was reported by another
group~\cite{checkelskyprl2009}, who proposed that the observed fluctuation
should not be categorized as conventional UCF.  They argue that the fluctuation
reflects the hybridization between the surface state and the bulk state confined
at surface.  A possible reason for the difference between their interpretation
and ours may arise from the difference in the thickness of the Bi$_{2}$Se$_{3}$
sample used; we use an ultra-thin film whereas they use a macroscopic crystal.
In the thin film, the bulk state confined to the surface may decrease, and
therefore, the effect of hybridization is likely to be reduced. Although we
currently do not know the definite reason for the difference between the two
results, it should be noted that the present analysis based on UCF is as
expected theoretically and is also consistent with the WAL effect.

\subsection{Contribution from the surface states}
Although we have reported the quantum transport phenomena of
Bi$_{2}$Se$_{3}$ thin film, within the experimental results presented here, it
is not easy to clarify to what extent the electronic properties as TI materials
are relevant. On the other hand, even in our experiment, there are a few
signatures that the properties as TI materials play some roles in the quantum
transport phenomena. One signature is that the surface states and the bulk
conduction states are at the Fermi level as seen in the ARPES result of
Fig. 1(a). This may suggest that electrons flow through the surface states and
the bulk conduction states.  The second one is based on the theoretical
report~\cite{Luprb2011} that a fitting parameter $\alpha$ of the HLN model can
be nearly equal to $-0.3$ if the surface states cause the WAL effect and the
bulk conduction states give rise to the weak localization effect. This is
consistent with our observation [see Fig. 4].  The above
discussion supports that the measured quantum transport phenomena is originated
from not only the bulk conduction states but also the surface states.  Further
experimental efforts, such as the electrical depletion of the bulk states
\cite{chenprb2011,Steinbergprb2011} or the carrier doping
\cite{Zhangnatcommun2011} are necessary to systematically detect phenomena
originated only from the Dirac electrons in the surface states of topological
insulators.

\section{Conclusion}
In summary, in the longitudinal resistance of a Bi$_{2}$Se$_{3}$ narrow Hall bar
sample, a dip near 0 T and a fluctuation as a sweep of field were observed.
This dip originates from the WAL effect, and we derived the coherence length by
using the HLN model.  By analyzing the temperature dependence
of the coherence lengths, it is confirmed
that conduction electrons move in a quasi-one-dimensional system.  The coherence length
calculated from the fluctuation on the assumption of UCF is consistent with that
calculated from the WAL effect based on the HLN model. The
conductance fluctuation as a function of temperature is consistent with the
prediction of the UCF theory.

\section{Acknowledgment}
This work is partially supported by the JSPS Funding Program for Next Generation
World-Leading Researchers.

\bibliography{ref.bib}

\begin{thebibliography}{42}
\bibitem [{att()}]{attention1}%
  \BibitemOpen
  \href@noop {} {}\bibinfo {note} {In the analysis of the coherence length, we
  do not use the data taken at the lowest system temperature (2 K), as we are
  not sure that the electron temperature reached down to this
  temperature.}\BibitemShut {Stop}%
\end{thebibliography}

\end{document}